\documentstyle[12pt]{article}
\newcommand{\eq}[1]{eq.(\ref{#1})}
\newcommand{\Eq}[1]{Eq.(\ref{#1})}
\newcommand{\ben}{\begin{equation}}
\newcommand{\een}{\end{equation}}
\newcommand{\bea}{\begin{eqnarray}}
\newcommand{\eea}{\end{eqnarray}}
\newcommand{\bear}{\begin{array}}
\newcommand{\enar}{\end{array}}
\newcommand{\bdm}{\begin{displaymath}}
\newcommand{\edm}{\end{displaymath}}
\newcommand{\nn}{\nonumber \\ }
\newcommand{\binomial}[2]{\left (\begin{array}{c} {#1}\\ {#2} \end{array}
\right )}

\newcommand{\pa}{\partial}

\newcommand{\Z}{\mbox{$Z\hspace{-2mm}Z$}}

\newcommand{\br}{\langle}
\newcommand{\kt}{\rangle}
\newcommand{\bra}[1]{\langle {#1}|}
\newcommand{\ket}[1]{|{#1}\rangle}

\newcommand{\dtp}[1]{\frac{d{#1}}{2\pi i}}

\newcommand{\vs}{\vspace}

\newcommand{\NP}[1]{Nucl.\ Phys.\ {\bf #1}}
\newcommand{\PL}[1]{Phys.\ Lett.\ {\bf #1}}
\newcommand{\CMP}[1]{Commun.\ Math.\ Phys.\ {\bf #1}}

\newcommand{\SJNP}[1]{Sov. J. Nucl. Phys.\ {\bf #1}}

\begin{document}

\topmargin -5mm
\oddsidemargin 5mm

\begin{titlepage}
\setcounter{page}{0}
\begin{flushright}
NBI-HE-95-19\\
hep-th/9506180\\
Revised version August 1995\\
\end{flushright}

\vs{8mm}
\begin{center}
{\Large HAMILTONIAN REDUCTION OF $SL(2)$-THEORIES}\\[.2cm]
{\Large AT THE LEVEL OF CORRELATORS}

\vs{8mm}
{\large Jens Lyng Petersen}\footnote{e-mail address:
 jenslyng@nbivax.nbi.dk},
{\large J{\o}rgen Rasmussen}
\footnote{e-mail address: jrasmussen@nbivax.nbi.dk}
{\large and Ming Yu}\footnote{e-mail address:
 yuming@nbivax.nbi.dk. Address after 1st March 1996:
Inst. of Theor. Phys., Academia Sinica, Beijing,
Peoples Republic of China}
\\[.2cm]
{\em The Niels Bohr Institute, Blegdamsvej 17, DK-2100 Copenhagen \O,
Denmark}\\[.5cm]

\end{center}

\vs{8mm}
\centerline{{\bf{Abstract}}}

Since the work of Bershadsky and Ooguri and Feigin and Frenkel
it is well known that correlators
of $SL(2)$ current algebra for admissible representations should reduce to
correlators for conformal minimal models. A precise proposal for this relation
has been given at the level of correlators: When $SL(2)$ primary fields are
expressed as $\phi_j(z_n,x_n)$ with $x_n$ being a variable to keep track
of the $SL(2)$ representation multiplet (possibly infinitely dimensional for
admissible representations), then the minimal
model correlator is supposed to be obtained simply by putting all $x_n=z_n$.
Although strong support for this has been presented, to the best of our
understanding a direct, simple proof seems to be missing so in this paper
we present one based on the free field Wakimoto construction and our previous
study of that in the present context. We further verify that the explicit
$SL(2)$ correlators we have published in a recent preprint reduce in the
above way, up to a constant
which we also calculate. We further discuss the relation to more standard
formulations of hamiltonian reduction.
\end{titlepage}
\newpage
\renewcommand{\thefootnote}{\arabic{footnote}}
\setcounter{footnote}{0}
\section{Introduction}
The relation between the $SL(2)$ current algebra and the Virasoro algebra via
hamiltonian reduction is well known \cite{Po}. In particular Bershadsky and
Ooguri \cite{BO} used the powerful BRST formalism for the reduction to
establish
equivalence between on the one hand $\widehat{SL}(2)_k$ WZNW theory after
reduction, and on the
other hand conformal minimal theory labelled by $(p,q)$, provided $k+2=p/q$,
and admissible representations were used in the WZNW case
(see also Feigin and Frenkel \cite{FF}). This equivalence was
discussed in those references at the level of the algebra and of the BRST
cohomology of physical states. A particularly simple and remarkable realization
of these ideas has been discussed by Furlan, Ganchev, Paunov and Petkova
\cite{FGPP} at the level of N-point conformal blocks
on the sphere. The formulation is in terms of primary fields of the affine
algebra in a formalism where they depend on two points \cite{FZ,ATY,FGPP,An}.
We use the notation of \cite{PRY}. Thus the algebra is given by
\bea
J^+(z)J^-(w)&=&\frac{2}{z-w}J^3(w)+\frac{k}{(z-w)^2}\nn
J^3(z)J^\pm(w)&=&\pm\frac{1}{z-w}J^\pm(w)\nn
J^3(z)J^3(w)&=&\frac{k/2}{(z-w)^2}
\eea
A primary field, $\phi_j(w,x)$, belonging to a representation labelled by
spin $j$, satisfies the following OPE's
\ben
J^a(z)\phi_j(w,x)=\frac{1}{z-w}{[}J_0^a,\phi_j(w,x){]}
\een
where the $SL(2)$ representation is provided by the differential operators
\bea
{[}J_n^a, \phi_j(z,x){]}&=&z^nD_x^a \phi_j(z,x)\nn
D_x^+&=&-x^2\pa_x+2xj\nn
D_x^3&=&-x\pa_x+j\nn
D_x^-&=&\pa_x
\eea
The statement in ref.\cite{FGPP} and under investigation in the present
paper is that correlators
of such primary fields should reduce to corresponding ones for a particular
minimal model in the limit where all $x$'s are put equal to the corresponding
$z$ values. We shall come back to a more precise statement which may also
be formulated for conformal blocks. The authors of ref.\cite{FGPP}
construct a solution of the Knizhnik-Zamolodchikov equations \cite{KZ} such
that
the $x_i$ dependence is described as a power series in $(x_i-z_i)$, and by
construction that solution is selected for which the boundary condition is,
that $x_i-z_i=0$ reproduces the corresponding minimal model correlator. The
expansion coefficients are given in terms of recurrence relations\footnote{
In fact it appears that the sums may be explicitly performed, V.B. Petkova,
private communication.}. To make sure that this solution of the KZ equations
really generates the WZNW correlator (up to normalization) a study is performed
in \cite{FGPP} of the null vector decoupling that follows from that solution,
and whether that
is as expected for a WZNW correlator. Although this was checked in many
examples
no explicit general proof was provided. The relation between null vector
decoupling in WZNW correlators and minimal model Virasoro ones was discussed
for example in ref.\cite{GP}. Thus, while there is very strong evidence for the
validity of the assertion, at least a simple and straightforward explicit proof
is not available. This is what we intend to supply in the present paper.

In ref.\cite{PRY} we have recently shown how to evaluate affine model conformal
blocks within the framework of the free field Wakimoto construction \cite{Wak}.
In particular
we developed a technique based on fractional calculus for how to deal with the
``second" screening charge written down in ref.\cite{BO} which involves
fractional powers of free fields and which is crucial for being able to deal
with the case of admissible representations. Very similar techniques (without
the present applications) have also been worked out by Andreev \cite{An}.

In sect. 2 we provide
a simple direct proof of the reduction between WZNW correlators and minimal
model ones as described above. However, we find that the affine
model conformal blocks differ from the minimal model ones by a normalization
factor which we evaluate and sometimes find to be zero or infinity,
making the procedure
of ref.\cite{FGPP} singular in those cases. Also we find that the result
obtains
in general just by putting all $x_i$'s {\em proportional} to all $z_i$'s
independent
of the factor of proportionality. This is reasonable because such a
proportionality constant would depend on normalizations of the currents.
In sect. 3 we compare with standard forms of hamiltonian reduction, and
in sect. 4 we explicitly verify that the correlator we wrote down in
ref.\cite{PRY} satisfies the theorem proven in this paper.

\section{Proof of the reduction at the correlator level}
The free field Wakimoto realization \cite{Wak} is obtained in terms of free
bosonic fields
of dimensions $(1,0)$ and of a scalar field which we take to have the
following contractions
\ben
\varphi(z)\varphi(w)=\log(z-w), \ \ \ \beta(z)\gamma(w)=\frac{1}{z-w}
\een
We only consider one chirality of the fields. The $\widehat{SL}(2)_k$ affine
currents may then be represented as
\bea
J^+(z)&=&\beta(z)\nn
J^3(z)&=&-:\gamma\beta:(z)-\sqrt{t/2}\pa\varphi(z)\nn
J^-(z)&=&-:\gamma^2\beta:(z)+k\pa\gamma(z)-\sqrt{2t}\gamma\pa\varphi(z)\nn
t&\equiv&k+2\neq 0
\label{wakimoto}
\eea
The Sugawara energy momentum tensor is obtained as
\ben
T(z)=:\beta\pa\gamma:(z)+\frac{1}{2}:\pa\varphi\pa\varphi:(z)+
\frac{1}{\sqrt{2t}}\pa^2\varphi(z)
\een
with central charge
$$c=\frac{3k}{k+2}$$
The primary field is \cite{PRY}
\bea
\phi_j(z,x)&=&(1+\gamma(z)x)^{2j}\phi_j^\varphi(z)\nn
\phi_j^\varphi(z)&=&:e^{-j\sqrt{2/t}\varphi(z)}:
\label{pfd}
\eea
where in general one should asymptotically expand $(1+\gamma(z)x)^{2j}$ as
\ben
(1+\gamma(z)x)_{(\alpha)}^{2j}=
\sum_{n\in \Z}\binomial{2j}{n+\alpha}(\gamma(z)x)^{n+\alpha}
\een
Here the choice of the parameter $\alpha$ depends on the monodromy conditions
of the primary field  $\phi_j(z,x)$ around contours in $x$-space, and those in
turn depend on the other fields present in the correlator, as is further
discussed in \cite{PRY}.

The primary field defined in \eq{pfd} may also be written as
\bea
\phi_j(z,x)
&=&e^{x\partial_y}\phi_j(z,y)|_{y=0}\nn
&=&e^{xD_y^-}\phi_j(z,y)|_{y=0}\nn
&=&e^{xJ^-_0}\phi_j(z,0)e^{-xJ^-_0}\nn
&=&e^{xJ^-_0}:e^{-j\sqrt{2/t}\varphi(z)}:e^{-xJ^-_0}
\label{pfd2}
\eea
Here there is a subtlety in that
the way the exponential function $e^{xJ^-_0}$ should be expanded \cite{PRY},
must also respect the monodromy conditions in the $x$ variables. We shall come
back to this subtlety presently. The two screening charges are
\bea
S_{k_\pm}(w)&=&\beta^{-k_\pm}(w)S^\varphi_\pm(w)\nn
S^\varphi_\pm(w)&=&:e^{-k_\pm\sqrt{2/t}\varphi(w)}:\nn
k_+&=&-1\nn
k_-&=&t
\label{screen}
\eea
A general conformal block (on the sphere) in the affine theory is then given by
\ben
W_N=\bra{j_N}R \phi_{j_{N-1}}(z_{N-1},x_{N-1})
...\phi_{j_{n}}(z_{n},x_n)...
\phi_{j_{2}}(z_2,x_2)\prod_i \oint \dtp{w_i}S_{k_i}(w_i)\ket{j_1}
\label{wn}
\een
where $R$ stands for radial ordering of the fields.
Different choices of integration contours for the screening charges define
different intertwining chiral vertex operators \cite{F,BF}, and different
conformal blocks.

The relation to minimal models is obtained by writing \cite{BO}
\bea
2j_{r,s}+1&=&r-st\nn
t&=&k+2=p/q \nn
\alpha_+&=& \sqrt{\frac{2}{t}}=-2/\alpha_- \nn
\alpha_{r,s+1}&=&-j_{r,s}\sqrt{\frac{2}{t}}
=\frac{1}{2}((1-r)\alpha_+-s\alpha_-)\nn
2\alpha_0&=&\alpha_++\alpha_-\nn
h_{r,s+1}&=& \frac{j_{r,s}(j_{r,s}+1)}{t}-j_{r,s}
=\frac{1}{2}\alpha_{r,s+1}(\alpha_{r,s+1}-2\alpha_0)\nn
\phi_{r,s+1}(z)&=& :e^{\alpha_{r,s+1}\varphi(z)}:
=\phi_{j_{r,s}}(z)\nn
V_{\alpha_{\pm}}(w)&=& :e^{\alpha_{\pm}\varphi(w)}:
=S_\pm^{\varphi}(w)
\eea
It is now clear that if one truncates the $\beta$-dependence of the screening
currents and the $\gamma$-dependent factor in the primary fields, then the
minimal model correlators are obtained \cite{FGPP}.
This is true despite the fact
that the two theories, the WZNW-model and the minimal model, have different
background charges for the $\varphi$-field:  namely
$-\alpha_+=-\sqrt{\frac{2}{t}}$ for the WZNW model and
$-2\alpha_0=-\sqrt{\frac{2}{t}}+\sqrt{2t}$ for the minimal models.
However, this difference is of
no consequence in the practical evaluation of the free field correlators since
in both cases suitable dual bra-states are used to absorb that background
charge
\cite{PRY}. Using \eq{pfd2} we may further write
\bea
\phi_j(z,x z)&=&e^{zx J^-_0}:e^{-j\sqrt{2/t}\varphi(z)}:
e^{-zx J^-_0}\nn
&=&e^{x zD^-_y}\phi_j(z,y)|_{y=0}\nn
&=&e^{x J^-_1}:e^{-j\sqrt{2/t}\varphi(z)}:e^{-x J^-_1}
\label{pfd3}
\eea
Consider the following conformal blocks
\ben
W_N=\bra{j_N}R \phi_{j_{N-1}}(z_{N-1},x z_{N-1})
...\phi_{j_{n}}(z_{n},x z_n)...
\phi_{j_{2}}(z_2,x z_2)\prod_i \oint \dtp{w_i}S_{k_i}(w_i)\ket{j_1}
\een
Substituting \eq{pfd3} we may rewrite this as
\ben
W_N=\bra{j_N}e^{x J^-_1}R \phi_{j_{N-1}}(z_{N-1},0)
...\phi_{j_{n}}(z_{n},0)...\phi_{j_{2}}(z_2,0)
\prod_i \oint \dtp{w_i}S_{k_i}(w_i)\ket{j_1}
\label{wn1}
\een
At this point however, there is a subtlety as to how adjacent exponentials
$$e^{\pm x J^-_1}$$
should be removed, and
we should examine how these exponentials are defined. Indeed as discussed
in \cite{PRY} the expansions of exponentials and other functions involving
the $\beta$- and $\gamma$-fields, depend on which monodromy the problem at hand
requires one to select. All these subtleties are dealt with using the following
two lemmas:

{\bf Lemma 1:}
If the fractional part in powers of $x$ is $\alpha $, then we can expand
the last expression in \eq{pfd2}
\ben
e^{xJ^-_0}:e^{-j\sqrt{2/t}\varphi(z)}:e^{-xJ^-_0}=
\sum_{n\in \Z}\frac{(xJ^-_0)^{\alpha-\beta+n}}{(\alpha-\beta +n)!}
:e^{-j\sqrt{2/t}\varphi(z)}:
\sum_{m\in \Z}\frac{(-xJ^-_0)^{\beta+m}}{(\beta +m )!}
\label{epd}
\een
for arbitrary complex number $\beta$.

{\bf Lemma 2:}
\ben
1=
\sum_{n\in \Z}\frac{(xJ^-_0)^{\alpha+n}}{(\alpha +n) !}
\sum_{m\in\Z}\frac{(-xJ_0^-)^{-\alpha +m}}{(-\alpha +m)!}
=e^{xJ^-_0}e^{-xJ^-_0}
\een
Before proving these lemmas we make the following remarks:
Define
\bea
\phi^{[n]}_j(z,0)
&=&[J^-_0,\phi^{[n-1]}_j(z,0)]\nn
\phi^{[0]}_j(z,0)&=& \phi_j(z,0)
\label{comm}
\eea
When $x$ is integrally powered, it is clear that we can expand $\phi_j(z,x)$ as
\bea
&&e^{xJ^-_0}:e^{-j\sqrt{2/t}\varphi(z)}:e^{-xJ^-_0}\nn
&=&\sum_{n\geq 0}\frac{\phi^{[n]}_j(z,0)x^n}{n!}\nn
&=&\sum_{n\geq 0}\frac{(D_y^-)^n\phi_j(z,y)x^n}{n!}|_{y=0}\nn
&=&e^{xD_y^-}\phi_j(z,y)|_{y=0}\nn
&=&\phi_j(z,x)
\label{scomm}
\eea
However, when $x$ is fractionally powered, we can no longer Taylor expand
$\phi_j(z,x)$, and the definition for both $\phi^{[n]}_j(z,0)$ in \eq{comm}
and $(D_y^-)^n\phi_j(z,y)|_{y=0}$ in \eq{scomm} requires specification.
It is still possible to generalize \eq{comm} and \eq{scomm} by defining
\ben
\frac{\phi^{[N+\alpha+\beta]}_j(z,0)}{(N+\alpha+\beta)!}=
\sum_{{n+m=N \atop n,m\in \Z}}\frac{(J^-_0)^{\alpha +n}}{(\alpha +n)!}
:e^{-j\sqrt{2/t}\varphi(z)}:\frac{(-J_0^-)^{\beta+m}}{(\beta+m)!}
\label{comm1}
\een
Although it looks like that the r.h.s. of \eq{comm1} depends on both $\alpha$
and $\beta$, lemma 1 essentially means that it only depends on the combination
$\alpha +\beta$.

The fractional derivatives at the origin may also be considered as analytical
continuations of their integral counterparts. Now
$\phi_j(z,x)=(1+\gamma(z)x)^{2j}:e^{-j\sqrt{2/t}\varphi(z)}:$, so for
non-negative integer $n$ we have
\ben
\frac{(D^-_y)^n}{\Gamma(n+1)}\phi_j(z,y)|_{y=0}
=\binomial{2j}{n}\gamma^n(z):e^{-j\sqrt{2/t}\varphi(z)}:
\een
We can analytically continue the variable $n$ in the above equation from
integers to complex numbers. Therefore $n$ could be any fractional
number and we have
\ben
\frac{(D^-_y)^{n+\alpha}}{\Gamma(n+\alpha +1)}\phi_j(z,y)|_{y=0}=
\binomial{2j}{n+\alpha}\gamma^{n+\alpha}(z):e^{-j\sqrt{2/t}\varphi(z)}:
\een

{\em Proof of Lemma 1:}
\bea
&&e^{xJ^-_0}:e^{-j\sqrt{2/t}\varphi(z)}:e^{-xJ^-_0}\nn &=&
\sum_{n\in \Z}\frac{(xJ^-_0)^{\alpha-\beta+n}}{(\alpha-\beta +n)!}
:e^{-j\sqrt{2/t}\varphi(z)}:
\sum_{m\in \Z}\frac{(-xJ^-_0)^{\beta+m}}{(\beta +m )!}\nn &=&
\sum_{n\in \Z}\frac{(xJ^-_0)^{\alpha-\beta+n}}{(\alpha-\beta +n)!}
\sum_{m\in \Z}\frac{(-xJ^-_0 + xD^-_y)^{\beta+m}}{(\beta +m) !}
\phi(z,y)|_{y=0}\nn &=&
\sum_{N\in \Z}\sum_{m\in \Z}
\frac{(xJ^-_0)^{\alpha-\beta+N-m}(-xJ^-_0 + xD^-_y)^{\beta+m}}
{(\alpha-\beta +N-m)! (\beta +m )!}\phi(z,y)|_{y=0}\nn &=&
\sum_{N\in \Z}
\frac{(xD^-_y)^{\alpha +N}}
{(\alpha +N) !}\phi(z,y)|_{y=0}\nn&=&
\sum_{N\in \Z}
\binomial{2j}{N+\alpha}(\gamma(z)x)^{N+\alpha}:e^{-j\sqrt{2/t}\varphi(z)}:\nn
&=&\phi_j(z,x)
\eea
This proves lemma 1.

{\em Proof of Lemma 2:}
\bea
e^{xJ^-_0}e^{-xJ^-_0}&=&
\sum_{n\in \Z}\frac{(xJ^-_0)^{\alpha+n}}{(\alpha +n) !}
\sum_{m\in \Z}\frac{(-xJ^-_0)^{-\alpha+m}}{(-\alpha +m )!}\nn
&=&\sum_{N\in \Z}\sum_{n+m=N}\frac{(xJ^-_0)^{N}}{(\alpha +n) ! (-\alpha +m) !}
(-1)^{-\alpha+m}\nn
&=&\sum_{N\in \Z}(xJ^-_0)^{N}\delta_{N,0}\nn
&=& 1
\eea
This proves lemma 2. Thus the manipulations in the proof in \eq{wn1} are
justified.

We may now go back to \eq{wn1}. We see from that expression that the
exponential
$$e^{x J^-_1}$$
has to be expanded in powers offset from the integers by the amount
$$-\sum k_i=r-st\equiv \alpha$$
with $r$ and $s$ the number of screening charges of the first and second
kind respectively, so this
is the combined power of the $\beta$-factors. Since the $\gamma$-factors
from the primary fields decouple for $x_i=0$, this particular expansion of the
exponential is required. One may then work out that
\ben
\bra{j_N}e^{x J^-_1}=\bra{j_N}(1-x\gamma_1)^{k-2j_N}
\frac{\sin{[}\pi(k-2j_N{]}}
{\sin{[}\pi(k-2j_N-\alpha)\pi{]}}(-)^{-\alpha}
\een
Notice that for $\alpha$ integer the ratio of sine-factors disappears
\footnote{This ratio of sine-factors, erroneously was missing in our first
version of the preprint for the present paper.}.
This result is obtained  by writing (for $\alpha = r-st$)
\bdm
e^{xJ^-_1}=\sum_{n\in\Z}\frac{(xJ^-_1)^{n+\alpha}}{(n+\alpha)!}
\edm
and observing that for any power
\bdm
\bra{j_N}(J^-_1)^{n+\alpha}=\bra{j_N}\gamma_1^{n+\alpha}
\frac{\Gamma(2j_N-k+n+\alpha)}{\Gamma(2j_N-k)}
\edm
The proportionality to a power of $\gamma_1$ follows from the free field
realization and the properties of the vacuum as described in \cite{PRY}. The
value of the constant is obtained by consistency between $n$ and $n+1$
and by normalizing with the result for $n+\alpha=1$.

We may now continue the calculation and obtain
\bea
W_N&=&\bra{j_N}(1-x\gamma_1)^{k-2j_N}\frac{\sin{[}\pi(k-2j_N){]}}
{\sin{[}\pi(k-2j_N-\alpha){]}}(-)^{-\alpha}\nn
&&R \phi_{j_{N-1}}(z_{N-1},0)
...\phi_{j_{n}}(z_{n},0)...\phi_{j_{2}}(z_2,0)
\prod_i \oint \dtp{w_i}S_{k_i}(w_i)\ket{j_1}\nn
&=&C_N(\{j_m\},x)\bra{j_N}R \phi_{j_{N-1}}(z_{N-1},0)
...\phi_{j_{n}}(z_{n},0)...\phi_{j_{2}}(z_2,0)
\prod_i \oint \dtp{w_i}S_{k_i}^{\varphi}(w_i)\ket{j_1}\nn
&=&C_N(\{j_m\},x)W_N^{\varphi}
\eea
where $C_N(\{j_m\},x)$ is the normalization constant, and $W_N^{\varphi}$ is
exactly the free field expression for the minimal model correlator.
The point is that since
\bdm
\bra{j_N}e^{x J^-_1}=\bra{j_N}(1-x\gamma_1)^{k-2j_N}
\frac{\sin{[}\pi(k-2j_N{]}}
{\sin{[}\pi(k-2j_N-\alpha)\pi{]}}(-)^{-\alpha}
\edm
now contains the only $\gamma$-dependence of the correlator, all
$\beta$-dependence is effectively removed from the screening charges
since $\gamma_1$ contracts only with
$\beta_{-1}$ which is the constant ($w_i$-independent) mode. Thus
(cf. \cite{PRY})
\ben
C_N(\{j_m\},x)=\frac{\Gamma(k-2j_N+1)\sin{[}\pi(k-2j_N){]}}
{\Gamma(k-2j_N-r+st+1)\sin{[}\pi(k-2j_N-r+st){]}}
x^{r-st}(-)^{-r+st}
\een
For $x=1$ one may check from the 3-point function given in \cite{PRY}
that this is indeed the relative constant to the 3-point function of minimal
models as given by Felder \cite{F}.
This is the simple proof of the statement presented in the beginning of this
paper.

\section{Comparison with standard formulations of hamiltonian reduction}
Having proved the equivalence of the two apparently different kinds of
correlators, we now want to understand this equivalence from the point of view
of quantum hamiltonian reduction. We briefly review the background.
Setting the Kac-Moody current
\ben
J^+(z)=1
\label{cstr}
\een
in the the equation of motion derived from the $SL(2)$ WZNW theory, one
recovers the classical equation of motion for Liouville theory. In order
to implement
the constraint, \eq{cstr}, at the quantum level, one introduces a lagrangian
multiplier field, $A(z)$, and follows the standard procedure for
hamiltonian reduction \cite{BO}, where $A(z)$ is treated as a gauge field.
The final theory, after gauge fixing, involves Faddeev-Popov ghost
fields, which are supposed to cancel out unwanted degrees of freedom
in the original WZNW theory.
The BRST quantization has now become a standard approach to constrained
hamiltonian systems. As far as correlation functions on the sphere are
concerned, the BRST quantization is equivalent to imposing the
constraint \eq{cstr} on the correlators. Suppose one writes the correlation
function on the sphere as an operator insertion
\ben
\bra{0}\hat{O}\ket{0}
\label{crt}
\een
then for the constrained system satisfying \eq{cstr}, we have
\ben
\bra{0}\hat{O}(J^+(z)-1)\ket{0}=0
\label{crt2}
\een
\Eq{crt2} is equivalent to the following condition
\bea
J^+_n\ket{0}&=&0\ \ \ \ \ n\geq 0\nn
\bra{0}\hat{O}(J^+_{-n}-\delta_{n,1})&=&0\ \ \ \ \ n\geq 1
\label{cdt}
\eea
In order not to confuse the notations in this paper, $J^+(z)$ is always
considered to be a conformal spin 1 field to fit the WZNW theory, so that
one has the expansion
\ben
J^+(z)=\sum_{n\in\Z}J^+_n z^{-n-1}
\een
As usual, to fix $J^+(z)$ to be a constant value would require $J^+(z)$
be a scalar field. In other words, the energy momentum tensor must be improved
from the Sugawara construction by adding a term $\partial_z J^3(z)$. In
that context, one should rename $J^+_n \rightarrow J^+_{n+1}$.

\Eq{cdt} is called the physical state condition. In BRST quantization
the physical state space is the same as the BRST cohomology space
$Ker(Q)/Im(Q)$,
where $Q$ is the BRST charge defined by
\ben
Q=\oint \dtp{w} (J^+(w)-1)c(w)
\label{brs}
\een
Here $c(w)$ is a conformal spin 1 fermionic ghost field with respect to the
improved
energy momentum tensor. Its conjugate field $b(w)$ is the antighost field of
spin 0 satisfying
\ben
b(w)c(z)=\frac{1}{w-z}
\een

\Eq{cdt} is equivalent to the BRST condition, in which one requires that the
vacuum states
$\bra{0}$ and $\ket{0}$ be physical states, and $\hat{O}$ be a physical
operator which maps physical states into physical states. In other words
\ben
\bra{0}Q=[Q,\hat{O}]=Q\ket{0}=0
\een

Now consider the most general form for a class of conformal blocks in
$SL(2)$ WZNW theory, which are proportional to those in the Virasoro minimal
models. They can be written in the following form
\bea
&&\bra{j_{r_N,s_N}}F(J^-_1)R \phi_{j_{r_{N-1},s_{N-1}}}(z_{N-1},0)
...
\phi_{j_{r_2,s_2}}(z_2,0)\prod_i \oint
\dtp{w_i}S_{k_i}(w_i)\ket{j_{r_1,s_1}}\nn
&=&C\bra{h_{r_N,s_N+1}}R \phi_{r_{N-1},s_{N-1}+1}(z_{N-1})
...\phi_{r_2,s_2}(z_2)
\prod_i \oint \dtp{w_i}V_{\alpha_i}(w_i)
\ket{h_{r_1,s_1+1}}
\label{equiv2}
\eea
where the normalization constant $C$ is found as before to be (up to a phase)
\bea
C&=&\frac{\Gamma(k-2j_N+1)\sin{[}\pi(k-2j_N){]}}
{\Gamma(k-2j_N+\sum_i k_i+1)\sin{[}\pi(k-2j_N+\sum_ik_i){]}}
(\partial_y)^{-\sum_i k_i}F(y)|_{y=0}\nn
&=&\frac{\Gamma(k-2j_N+1)\sin{[}\pi(k-2j_N){]}}
{\Gamma(k-\sum_{i=1}^N j_i+1)\sin{[}\pi(k-\sum_{i=1}^N j_i){]}}
(\partial_y)^{-2j_N+\sum_{i=1}^{N} j_i}F(y)|_{y=0}\nn
\eea
In general $C$ depends on $t$ and the $j_i$'s. For some values of
$t$ and $j_i$'s, $C$ vanishes. Then the conformal blocks in the Virasoro
minimal models can only be obtained by dividing out $C$.
In other cases $C$ becomes infinity.
Then the conformal blocks for the Virasoro minimal models can
be either finite or zero, and in the latter case the relation between the WZNW
and the minimal conformal blocks is singular. Strictly speaking,
simply taking the limit $x_i \rightarrow z_i$ is not equivalent to quantum
hamiltonian reduction \eq{cstr}. Rather it is in accord with the constraint
\ben
J^+(w)=J^+_{-1}
\label{cstr2}
\een
To go to the minimal model we must further impose the condition
\ben
J^+_{-1}=1
\label{cstr3}
\een
To see this, let us consider the BRST charge for quantum hamiltonian reduction
\eq{cstr2}
\ben
\tilde{Q}=\oint \dtp{w} (J^+(w)-J^+_{-1})c(w)
\label{brs2}
\een
The physical state space now becomes the BRST cohomology space
$Ker(\tilde{Q})/Im(\tilde{Q})$. It is clear that
$\phi_j(z,0)$ commutes with $\tilde{Q}$, hence maps a physical state into
another physical state. Now consider the ket and the bra states.
Notice that the ket state $\ket{j_1}$ is a highest weight state and the
bra state $\bra{j_{r_N,s_N}}$ is a lowest weight state
\ben
J^+_n\ket{j_1}=\bra{j_{r_N,s_N}}J^+_{-n-1}=0, \ \ \ \ n\geq 0
\label{jpktbr}
\een
For the $b$, $c$ ghost fields, we have the following condition
\ben
c_n\ket{j_1}=b_{n+1}\ket{j_1}
=\bra{j_{r_N,s_N}}c_{-n-1}
=\bra{j_{r_N,s_N}}b_{-n}
=0, \ \ \ \ n\geq 0
\een
It can be verified that
with respect to the BRST charge $\tilde{Q}$ in \eq{brs2},
$\ket{j_1}$ is a physical state, and
the bra state $\bra{j_{r_N,s_N}}F(J^-_1)$ is a physical state  for
any arbitrary function $F(J^-_1)$. However, this extra degree of
freedom is removed if we further impose the condition \eq{cstr3},
which would fix the function  $F(J^-_1)$ uniquely, and
we recover exactly the conformal blocks in the Virasoro minimal models
\ben
\bra{j_{r_N,s_N}}F(J^-_1)=
\bra{j_{r_N,s_N}}e^{-\gamma_1}
=\bra{j_{r_N,s_N}}\sum_{n\in \Z}
\frac{\Gamma(2j_N-k)}{\Gamma(2j_N-k+n+\alpha)\Gamma(n+\alpha+1)}
(J^-_1)^{n+\alpha}
\een
where $\gamma_n$ is conjugate to $J^+_{-n}$
\ben
[J^+_{-n}, \gamma_m]=\delta_{n,m}
\een
If we were to use $\phi_j(z,z)$ to represent a primary field in the hamiltonian
reduced system (strictly speaking, $\phi_j(z,z)$ does not transform as a
primary field for the Virasoro algebra in the reduced system), then we
should normalize the correlation function by
dividing out the normalization constant $C$. Then in the limit $C$ goes
to zero, the conformal block in the reduced system would remain finite.

In conclusion, the constraint $J^+(z)=1$ completely freezes the degrees
of freedom of the $J^+(z)$ field. However we could proceed in two steps in
putting the constraint on the correlation functions. First set
$J^+(z)=J^+_{-1}$ and then let $J^+_{-1}=1$. The first step would result in
a class of correlation functions which are proportional to that of the
completely constrained system, like the ones considered in the previous
section. However, the remaining degrees of freedom
of the $J^+_{-1}$ mode is reflected by the arbitrariness of the
proportionality. If we normalize the correlation function by dividing
out the normalization constant, which is equivalent to setting $J^+_{-1}=1$,
then we recover the corresponding
correlators in the completely reduced system.
\section{Analysis of the explicit correlators}
Finally we want to verify explicitly that the conformal blocks for the
WZNW model evaluated in ref.\cite{PRY} satisfy the results of sect. 2.
To this end we consider the interpolating correlator
\ben
\bra{j_N}e^{-J^-_1x_N}\prod_{\ell =2}^{N-1}\phi_{j_{N-1}}(z_{N-1},x_{N-1})...
\phi_{j_2}(z_2,x_2)\prod_{i=1}^M\oint\dtp{w_i}S_{k_i}(w_i)\ket{j_1}\equiv
\bra{j_N}{\cal O}\ket{j_1}
\een
with
\bea
x_\ell &=&z_\ell x, \ \ \ell=1,...,N-1\nn
x_N&=&x-1
\eea
Thus for $x=1$ we get the WZNW model with all $x_i$'s put equal to the $z_i$'s.
For $x=0$ we should get the minimal model correlator. We wish to show that when
this interpolating correlator is evaluated according to ref.\cite{PRY}
then indeed it is independent of $x$. Using ref.\cite{PRY} we find
\ben
\br{\cal O}\kt=W_BW_N^\varphi F
\een
where
\bea
B(w)&=&\sum_{\ell =1}^{N-1}\frac{x_\ell/u_\ell}{w-z_\ell}-x_N/u_N\nn
W_B&=&\prod_{i=1}^MB(w_i)^{-k_i}\nn
F&=&\Gamma(k-2j_N+1)u_N^{k-2j_N-1}e^{1/u_N}\prod_{\ell=2}^{N-1}
\Gamma(2j_\ell +1)u_\ell^{2j_\ell -1}e^{1/u_\ell} \nn
W^\varphi_N&=&\prod_{m<n}(z_m-z_n)^{2j_mj_n/t}\prod_{i=1}^M\prod_{m=1}^{N-1}
(w_i-z_m)^{2k_ij_m/t}\prod_{i<j<M}(w_i-w_j)^{2k_ik_j/t}
\eea
Here we used that
\bea
\bra{j_N}e^{-x_NJ^-_1}&=&\bra{j_N}(1+x_N\gamma_1)^{k-2j_N}\nn
&=&\lim_{z'\rightarrow\infty}\bra{j_N}(1+x_N\gamma(z'))^{k-2j_N}
\eea
This may seem in contradiction to the expansion we used above in sect. 2, since
the ratio of sine-functions we had there is absent now. So let us explain
the reason for this subtlety. The point is that for $x\neq 0$ the
fractional powers of the $\beta$'s in the correlator is balanced by the
fractional powers of the $\gamma$'s in the primary fields. Thus the exponential
in the present case has to be expanded in {\em integral} powers.
This is in contrast
to the situation in sect. 2 where the $\gamma$-dependence of the primary
fields were suppressed since they involved the case $x_i=0$. One may then ask
how it is that this ratio of sine-factors is recovered in the present context.
Indeed, naively putting $x=0$ in the expression for the $B$ factors renders
the $u$-integrals trivial and we recover erroneously the factor $C_N(\{j_m\})$
without the ratio of sine-factors. However a more careful analysis of the cut
structure and of the integration contours shows that in fact this ratio arises
again when care is exercised. To see this, it is convenient to change variable
form $u_N$ to $U_N\equiv 1/u_N$. As a function of $U_N$ the integrand has a
branch point singularity at $U_N=0,\infty,U_N(x)$ with
$$U_N(x)=\sum_{\ell =1}^{N-1}\frac{xz_\ell/u_\ell}{(w-z_\ell)(x-1)}$$
Now we should remember that all the $u$-integrals are to be taken along
{\em small} circles surrounding the origin \cite{PRY}, and we take them all to
have the same radius (which is taken to zero at the end). The $U_N$ contour
is therefore some {\em large } circle. However, for $x$ close to $1$,
we see that
$U_N(x)$ will lie {\em outside} this circle, which we therefore may deform
along a contour running above and below the negative real axis. As $x$ is
decreased from $1$, the branch point at $U_N(x)$ moves closer to the original
integration circle for $U_N$ and eventually crosses it. Therefore,
by analytic continuation this circle has to be
deformed so as to keep the singularity always outside. A convenient way of
doing  that is precisely by deforming it immediately to wrap around the
negative
axis. But then the contributions form above and below the axis will have
different phases, and it is not difficult to see that these exactly reproduce
the seemingly missing ratio of sine-factors.

Consider now
\bea
G^-(w)&=&\br J^-(w){\cal O}\kt\nn
&=&\left \{-\sum_{i,j}B(w)\frac{D_{B_i}D_{B_j}}{(w-w_i)(w-w_j)}
+(t-2)\sum_i\frac{D_{B_i}}{(w-w_i)^2}\right.\nn
&-&\left.2\sum_{i,j}\frac{k_iD_{B_j}}{(w-w_i)(w-w_j)} -
2\sum_{m,j}\frac{j_mD_{B_j}}{(w-z_m)(w-w_j)}\right\}W_BW^\varphi_NF
\eea
This function has simple poles as a function of $w$. It is a rather simple
matter to evaluate the pole residues along the lines described in
ref.\cite{PRY}. The result is
\bea
\oint_{z_m} \dtp{w}G^-(w)w&=&z_m\pa_{x_m}W_BFW^\varphi_N\nn
\oint_\infty\dtp{w}G^-(w)w&=&\pa_{x_N}W_BFW^\varphi_N\nn
\oint_{w_i} \dtp{w}G^-(w)w&=&t\pa_{w_j}
\left(w_j\frac{W_N^\varphi W_B F}{B(w_j)}\right )
\eea
After integration over the $w_i$'s we see that we precisely produce the total
derivative of the original correlator with respect to $x$:
\ben
\sum_mz_m\pa_{x_m}W_B+\pa_{x_N}W_B=\pa_xW_B
\een
This expression will vanish since this
merely is the condition that the sum of pole residues vanishes (when the pole
at infinity is included as it is here).\\[.5cm]
{\bf Acknowledgement}\\[.3cm]
We are grateful to A.Ch. Ganchev and V.B. Petkova for explaining several
details of the work in
refs.\cite{FGPP,GP} to us, and for other illuminating remarks, and we are
grateful to O. Andreev for having pointed out his work, \cite{An}, to us.
M.Y. thanks the Danish research Academy for financial support.


\begin{thebibliography}{99}
\bibitem{Po}A.M. Polyakov, in {\em Physics and Mathematics of Strings,} Eds.
L. Brink, D. Friedan and A.M. Polyakov (World Scientific, 1990)
\bibitem{BO} M. Bershadsky and H. Ooguri, \CMP{126} (1989) 49
\bibitem{FF} B.L. Feigin and E. Frenkel, Lett. Math. Phys. {\bf 19} (1990) 307
\bibitem{FGPP}P. Furlan, A.Ch. Ganchev, R. Paunov and V.B. Petkova,
\PL{B267} (1991) 63;\\
P. Furlan, A.Ch. Ganchev, R. Paunov and V.B. Petkova,
\NP{B394} (1993) 665;\\
A.Ch. Ganchev and V.B. Petkova, \PL{B293} (1992) 56
\bibitem{FZ} V.A. Fateev and A.B. Zamolodchikov, \SJNP{43} (1986) 657
\bibitem{ATY} H. Awata, A. Tsuchiya and Y. Yamada, \NP{B365} (1991) 680;\\
H. Awata, KEK-TH-310/KEK Preprint 91-189, preprint
\bibitem{An} O. Andreev, HUB-IEB-94/9 hep-th/9407180 preprint, and
Landau-95-TMP-1/hep-th/9504082, preprint
\bibitem{PRY} J.L. Petersen, J. Rasmussen and M. Yu,
NBI-HE-95-16/hep-th/9504127, preprint
\bibitem{KZ} V. Knizhnik and A. Zamolodchikov, \NP{B247} (1984) 83
\bibitem{GP} A.Ch. Ganchev and V.B. Petkova, \PL{B318} (1993) 74
\bibitem{Wak} M. Wakimoto, \CMP{104} (1986) 605
\bibitem{F} G. Felder, \NP{B317} (1989) 215 {[}Erratum: {\bf B324} (1989)
548{]}
\bibitem{BF} D. Bernard and G. Felder, \CMP{127} (1990) 145
\end{thebibliography}
\end{document}